\documentclass[twocolumn,letterpaper,prb,showpacs,amsmath]{revtex4}

\usepackage{graphicx}

\begin{document}

\title{Quantum mechanical picture of the coupling between interlayer electronic excitations 
and infrared active phonons in bilayer cuprate superconductors}

\author{Ji\v{r}\'{\i} Va{\v s}{\'a}tko}
\author{Dominik Munzar}
\affiliation{Department of Condensed Matter Physics, Faculty of Science,
and Central European Institute of Technology, 
Masaryk University, Kotl\'a\v{r}sk\'a 2, 61137 Brno, Czech Republic}

\begin{abstract}
The formula frequently used to describe the $c$-axis infrared response 
of the coupled electron-phonon system of bilayer cuprate superconductors
and providing important insights into the physics of these materials
has been originally obtained 
at the level of the phenomenological multilayer model. 
Here we derive it using diagrammatic perturbation theory. 
\end{abstract}

\date{\today}


\maketitle 

The phase diagram of underdoped high-$T_{\mathrm{c}}$ cuprate superconductors 
is divided into three regions: 
the superconducting one below $T_{\mathrm{c}}$, 
the pseudogap phase in the temperature range $T_{\mathrm{c}}<T<T^{*}$, 
where $T^{*}$ is the onset temperature of the (high energy) pseudogap, 
and the anomalous normal state above $T^{*}$.\cite{Timusk:1999:RPP,Norman:2005:AP} 
There is an interesting and important class of physical properties, 
that begin to develop/whose temperature dependence changes its character 
at an onset temperature much lower than $T^{*}$
but at the same time considerably higher than $T_{\mathrm{c}}$.\cite{Timusk1:2003:SSC}    
These include, e.g., 
the NMR relaxation rate ($1/T_{1}T$),\cite{Warren:1989:PRL} 
amplitude of the transverse plasma mode (also referred to as the $400{\rm\,cm}^{-1}$ mode) 
and some phonon structures 
in the $c$-axis infrared conductivity of YBa$_{2}$Cu$_{3}$O$_{7-\delta}$ 
(Y-123)~~\cite{Homes:1993:PRL,Homes:1995:PhysicaC,Schuetzman:1995:PRB,Bernhard:2000:PRB,Timusk2:2003:SSC,Dubroka:2011:PRL},  
in-plane scattering rate and in-plane conductivity 
below ca $700{\rm\,cm}^{-1}$~~\cite{Timusk1:2003:SSC,Dubroka:2011:PRL},    
amplitude of the neutron resonance\cite{Hinkov:2007:NaturePhys},  
Nernst signal\cite{Wang:2006:PRB},  
diamagnetic response\cite{Li:2010:PRB}, 
specific heat\cite{Tallon:2011:PRB}, 
some features of the excitation gap\cite{Okada:2011:PRB,Krasnov:2009:PRB}, 
spectral weight at the chemical potential\cite{Kondo:2011:NaturePhys}, 
magnetic field dependence of the in-plane resistivity\cite{Rullier:2011:PRB},   
THz conductivity\cite{Corson:1999:Nature,Bilbro:2011:NaturePhys}, and  
the quasiparticle interference\cite{Lee:2009:Science}. 
The values of the onset temperature range 
from ca $15{\rm\,K}$ above $T_{c}$ (THz conductivity)
to more than $100{\rm\,K}$ above $T_{c}$ (transverse plasmon and the phonon structures, Nernst). 
Some of the phenomena 
(increase of the THz conductivity and quasiparticle interference)  
are almost certainly related to superconducting fluctuations. 
Some other with a much higher onset temperature  
(e.g., the transverse plasmon and the phonon structures, Nernst),  
have also been attributed to a precursor superconducting state (PSC)
lacking the long range phase coherence. 
This exciting issue, however, 
is fairly controversial and no consensus has been reached thus far. 
There are two prerequisites for an ultimate understanding: 
(a) A detailed knowledge of both doping ($p$) and temperature dependence of the relevant quantities. 
(b) An understanding of the physical contents of these quantities, 
in particular, the relation to superconductivity. 

Dubroka {\it et al.}\cite{Dubroka:2011:PRL} have studied in detail  
the temperature dependence of the phonon structures in the infrared $c$-axis conductivity of Y-123
for a broad range of $p$ of $0.03<p<0.17$ 
and determined the $p$-dependence of the corresponding onset temperature $T^{\mathrm{ons}}$. 
The relation between the phonons and the electronic degrees of freedom has been understood 
in terms of the phenomenological, essentially classical,  
multilayer model\cite{Marel:1996:CzJP,Grueninger:2000:PRL},  
involving the local conductivity of the bilayer units (denoted here as $\sigma_{\mathrm{bl}}$) 
and that of the spacing layers ($\sigma_{\mathrm{int}}$),  
extended\cite{Munzar:1999:SSC,Dubroka:2004:PhysicaC} 
by including formulas for the local fields acting on the relevant ions 
and a feedback of the phonons. 
For a schematic representation of the electronic part of the model, 
see Fig.~1 (b) of Ref.~\onlinecite{Chaloupka:2009:PRB}.  
For a careful discussion of the role of the phonons, 
see Ref.~\onlinecite{Dubroka:2004:PhysicaC}.  
The model has allowed the authors of Ref.~\onlinecite{Dubroka:2011:PRL} to conclude      
that at $T^{\mathrm{ons}}$ 
the spectral weight of the low-energy component of $\sigma_{\mathrm{bl}}$ 
(the Drude peak in the analysis of Fig.~1 of Ref.~\onlinecite{Dubroka:2011:PRL})  
begins to increase, signalling an increase of coherence of the response. 
Several arguments have been advanced suggesting that   
this is due to the formation of a PSC.   
The importance of the topic of PSC and of other interesting fields of application 
of the multilayer model \cite{Boris:2003:PRL,Hirata:2010:PhysicaC,LaForge:2007:PRB,Marek:2011:JPCM} 
calls for developing a fully quantum mechanical description 
of the response of the coupled electron-phonon system,  
that would support the phenomenological approach.  
Chaloupka, Bernhard, and Munzar have already constructed 
(see Ref.~\onlinecite{Chaloupka:2009:PRB}, CBM in the following) 
a microscopic gauge-invariant theory of the $c$-axis infrared response 
of bilayer cuprate superconductors, such as Y-123, 
providing both a justification and limits of the electronic part of the multilayer model
and of a related earlier approach \cite{Shah:2001:PRB}.   
The phonons, however, have not yet been incorporated. 
Here we provide and extension of the theory of CBM, 
where they are included at the Green's function level. 
A similar approach has been recently used to explain the electric field dependence 
of the infrared spectra of bilayer graphene\cite{Cappelluti:2010:PRB}. 
The formula for the $c$-axis conductivity will be derived 
and it will be shown to coincide, 
for a natural choice of the electron-phonon coupling term,  
with that of the phenomenological model. 
The essential aspects of the theory are formulated in the points (i)-(viii).  

For the sake of simplicity 
we neglect the hopping through the spacing layer ($t'_{\perp}$ in the notation of CBM, 
$\sigma_{\mathrm{int}}=0$),  
focus, as usual, on the long-wavelength limit, ${\mathbf q}=0$,\cite{finiteq}
and limit ourselves to the case of one infrared active phonon. 

(i) {\it Relevant equations of CBM}. 
The (electronic) conductivity $\sigma_{\mathrm bl}$ 
of~the~bilayer unit is given by 
\begin{equation}
\sigma_{\mathrm{bl}}(\omega)=-i\omega\varepsilon_{0}\chi_{\mathrm{bl}}(\omega)=
{\kappa_{\mathrm{dia-bl}}+d_{\mathrm{bl}}\Pi_{jj}(\omega)\over i(\omega+i\delta)}\,, 
\label{eq:FromI1}
\end{equation}
where $\chi_{\mathrm{bl}}$ is the bilayer susceptibility,   
$\kappa_{\mathrm{dia-bl}}$ the term describing the diamagnetic contribution, 
$d_{\mathrm{bl}}$ the distance between the closely spaced planes,      
and $\Pi_{jj}(\omega)$ the current-current correlation function. 
The term $\kappa_{\mathrm{dia-bl}}$ is given by 
\begin{equation}
\kappa_{\mathrm{dia-bl}}=-{e^{2}d_{\mathrm{bl}}\over \hbar^{2}Na^{2}}
\sum_{{\mathbf k}s}t_{\perp}({\mathbf k})
\langle 
c_{B{\mathbf k}s}^{\dagger}
c_{B{\mathbf k}s}-
c_{A{\mathbf k}s}^{\dagger}
c_{A{\mathbf k}s}
\rangle\,,
\label{eq:FromI2}
\end{equation}
where $N$ is the number of surface unit cells per copper-oxygen plane, 
$a$ is the in-plane lattice parameter,   
the sum runs over all values of the in-plane quasimomentum ${\mathbf k}$ and spin $s$,  
$t_{\perp}({\mathbf k})$ is the intra-bilayer hopping parameter, and   
$c_{B{\mathbf k}s}^{\dagger}$, $c_{B{\mathbf k}s}$ 
($c_{A{\mathbf k}s}^{\dagger}$, $c_{A{\mathbf k}s}$) 
are the quasiparticle operators corresponding to the bonding (antibonding) band. 
The correlator $\Pi_{jj}$ is given by  
\begin{equation}
\Pi_{jj}(\omega)={i Na^{2}\over \hbar}\int_{-\infty}^{\infty}\,{\mathrm d}t\, 
e^{i\omega t}\langle 
[{\hat j}_{\mathrm{bl}}^{\mathrm{p}}(t),{\hat j}_{\mathrm{bl}}^{\mathrm{p}}(0)]
\rangle \theta(t)\,, 
\label{eq:FromI3}
\end{equation}
where 
${\hat j}_{\mathrm{bl}}^{\mathrm{p}}$ is the operator of the intrabilayer paramagnetic current density, 
\begin{equation}
{\hat j}_{\mathrm{bl}}^{\mathrm{p}}={ie\over \hbar Na^{2}}
\sum_{{\mathbf k}s}t_{\perp}({\mathbf k})
\left(
c_{B{\mathbf k}s}^{\dagger}c_{A{\mathbf k}s}- 
c_{A{\mathbf k}s}^{\dagger}c_{B{\mathbf k}s}
\right)\,.
\label{eq:FromI4}
\end{equation}   
In the absence of interlayer Coulomb interaction and any phonon contribution,  
the total (volume averaged) $c$-axis current density 
would be equal to $\sigma^{*}(\omega)E^{\mathrm{ext}}(\omega)$
with $\sigma^{*}(\omega)=(d_{\mathrm{bl}}/d)\sigma_{\mathrm{bl}}(\omega)$ 
and $E^{\mathrm{ext}}$ the external field,  
$d$ is the $c$-axis lattice parameter.  

(ii) {\it Response  
of a $c$-axis polarized infrared active phonon mode. }
In the absence of any electronic contribution, 
the current density can be expressed as    
$\sigma^{*}(\omega)E^{\mathrm{ext}}(\omega)$ with  
\begin{equation}
\sigma^{*}(\omega)=
\sigma_{\mathrm{ph}}(\omega)=-i\omega\varepsilon_{0}\chi=
{\kappa_{\mathrm{dia-ph}}+d\Lambda_{jj}(\omega)\over i(\omega+i\delta)}\,,
\label{eq:Fonon1}
\end{equation}
where $\chi$ is the phonon polarizability, 
$\kappa_{\mathrm{dia-ph}}$  the term describing the diamagnetic contribution, 
and $\Lambda_{jj}(\omega)$ the phonon current-current correlator. 
The term $\kappa_{\mathrm{dia-ph}}$ is equal to $-\varepsilon_{0}\omega_{P}^{2}$, 
where $\omega_{P}$ is the phonon plasma frequency, 
$\omega_{P}^{2}=1/(a^{2}d\varepsilon_{0})
\left(\sum_{\mu}[e_{\mu}({\mathbf e})_{\mu\,z}/\sqrt{m_{\mu}}] \right)^{2}$. 
Here $e_{\mu}$, $({\mathbf e})_{\mu\,z}$ and $m_{\mu}$ are 
the effective charge, the $c$-axis component of the polarization vector and the mass 
of the $\mu$-th ion.  
The correlator is given by 
\begin{equation}
\Lambda_{jj}(\omega)={i Na^{2}\over \hbar}\int_{-\infty}^{\infty}\,{\mathrm d}t\, 
e^{i\omega t}\langle 
[{\hat j}_{\mathrm{ph}}^{\mathrm{p}}(t),{\hat j}_{\mathrm{ph}}^{\mathrm{p}}(0)]\rangle \theta(t)\,. 
\label{eq:Fonon2}
\end{equation} 
Here ${\hat j}_{\mathrm{ph}}^{\mathrm{p}}$ is the phonon paramagnetic current-density operator, 
\begin{equation}
{\hat j}_{\mathrm{ph}}^{\mathrm{p}}={i\over a^{2}d}\sqrt{\hbar \omega_{0}\over 2}
\sum_{\mu}{e_{\mu}({\mathbf e})_{\mu\,z}\over \sqrt{m_{\mu}}}(a^{\dagger}-a)\,,  
\label{eq:Phonon3} 
\end{equation}
$\omega_{0}$ the phonon frequency, 
$a^{\dagger}$ and $a$ are the phonon creation and annihilation operators, respectively.  
The correlator can be further expressed in terms of the retarded phonon propagator $D(\omega)$:
\begin{equation}  
d\Lambda_{jj}=\varepsilon_{0}\omega_{0}^{2}\chi=
-{\varepsilon_{0}\omega_{P}^{2}\omega_{0}\over 2}D(\omega)\,. 
\label{eq:Phonon4} 
\end{equation}
In the noninteracting case, \\ 
$D(\omega)=D_{0}(\omega)=2\omega_{0}/(\omega^{2}-\omega_{0}^{2}+i\omega\delta)$. 

(iii) {\it Case of the electronic contribution coexisting with the phonon one.} 
The total current density is given by $\sigma^{*}(\omega)E^{\mathrm{ext}}$ with 
\begin{equation}
\sigma^{*}(\omega)=
{\kappa'_{\mathrm{dia-bl}}+\kappa_{\mathrm{dia-ph}}+d\,\Xi_{jj}(\omega)\over i(\omega+i\delta)}\,,
\label{eq:Elandph1}
\end{equation}
where $\kappa'_{\mathrm{dia-bl}}=(d_{\mathrm{bl}}/d)\kappa_{\mathrm{dia-bl}}$, 
$$
\Xi_{jj}(\omega)=
{i Na^{2}\over \hbar}\int_{-\infty}^{\infty}\,{\mathrm d}t\, 
e^{i\omega t} \times
$$
\begin{equation}
\times 
\langle 
[{\hat J}_{\mathrm{bl}}^{\mathrm{p}}(t)+{\hat j}_{\mathrm{ph}}^{\mathrm{p}}(t),
{\hat J}_{\mathrm{bl}}^{\mathrm{p}}(0)+{\hat j}_{\mathrm{ph}}^{\mathrm{p}}(0)]\rangle \theta(t) 
\label{eq:Elandph2}
\end{equation} 
and 
${\hat J}_{\mathrm{bl}}^{\mathrm{p}}=(d_{\mathrm{bl}}/d){\hat j}_{\mathrm{bl}}^{\mathrm{p}}$. 
In the absence of interlayer Coulomb interaction and electron-phonon interaction, 
the mixed correlators (involving both ${\hat J}_{\mathrm{bl}}^{\mathrm{p}}$ 
and ${\hat j}_{\mathrm{ph}}^{\mathrm{p}}$) 
would vanish and we would obtain simply 
$\sigma^{*}(\omega)=(d_{\mathrm{bl}}/d)\sigma_{\mathrm{bl}}(\omega)+\sigma_{\mathrm{ph}}(\omega)$ 
with $\sigma_{\mathrm{bl}}$ from Eq.~(\ref{eq:FromI1}) and $\sigma_{\mathrm{ph}}$ from Eq.~(\ref{eq:Fonon1}).
The interactions will be shown to lead 
both to a renormalization of $\sigma_{\mathrm{bl}}$ and $\sigma_{\mathrm{ph}}$
and to nonzero values of the mixed correlators. 

(iv) {\it Formulas used to describe the final state interactions.} 
Let us begin our considerations with the simple scheme of Fig.~1 (b) of CBM. 
The difference between the local conductivities $\sigma_{\mathrm{bl}}$ and $\sigma_{\mathrm{int}}$ 
leads to a difference between the current densities $j_{\mathrm{bl}}$ and $j_{\mathrm{int}}$
resulting in~a~charge redistribution between the CuO$_{2}$ planes, 
that can be characterized by~the excess surface charge density $\rho$ of the bottom plane
(the density of the upper plane is $-\rho$).   
The~electrostatic energy due to $\rho$ is given by $Na^{2}d_{\mathrm{bl}}\rho^{2}/(2\varepsilon_{0})$ 
and the corresponding contribution to~the~Hamiltonian by~Eq.~(25) of CBM, i.e.,   
\begin{equation}
H_{\mathrm{C}}={Na^{2}d_{\mathrm{bl}}\over 2\varepsilon_{0}}{\hat{\rho}}^{2}\,,
\label{eq:InterlayerCoulomb}
\end{equation}
where ${\hat{\rho}}$ is the density operator,  
that can be expressed in terms of the quasiparticle operators as 
\begin{equation}
{\hat \rho}={e\over 2Na^{2}}
\sum_{{\mathbf k}s}
\left(
c_{A{\mathbf k}s}^{\dagger}c_{B{\mathbf k}s}+ 
c_{B{\mathbf k}s}^{\dagger}c_{A{\mathbf k}s}
\right)\,.
\label{eq:Formulaforrho}
\end{equation}
Next we derive the formula for the coupling term $H_{\mathrm{e-p}}$ 
describing the interaction between the~surface charge densities of the planes $\rho$ and $-\rho$ and a phonon. 
We begin with the textbook formula for the energy $U$ of a polarized medium in a homogeneous electric field:
$U=-V \cdot P \cdot E$, 
where $V$ is the volume of the medium, $P$ the polarization and $E$ the field. 
In the present case, 
$V$ is the volume of $N$ unit cells, $V \rightarrow Na^{2}d$,  
$P$ is the polarization due to the phonon, $P\rightarrow {\hat P}$,  
\begin{equation}
{\hat P}=
{1\over a^{2}d}\sqrt{\hbar \over 2\omega_{0}}
\sum_{\mu}{e_{\mu}({\mathbf e})_{\mu\,z}\over \sqrt{m_{\mu}}}(a^{\dagger}+a)\,,
\label{eq:Phononpolarization} 
\end{equation}
and $E$ is the electric field due to $\rho$ and $-\rho$, 
$E\rightarrow {\hat \rho}/\varepsilon_{0}$.  
In addition, the product has to be multiplied by a factor denoted by $\xi$, 
representing the degree of overlap of $E$ and the microscopic dipoles 
($E$ is nonzero only in the bilayer unit); 
$\xi=1$($0$) for phonons associated with vibrations inside the bilayer unit 
(outside the bilayer unit) and $\xi\approx 1/2$ for phonons involving mainly    
the oxygens of the CuO$_{2}$ planes.     
By combining the expressions for $V$, $P$, and $E$ we obtain 
\begin{equation}
H_{\mathrm{e-p}}=-\xi{Na^{2}d\over\varepsilon_{0}}{\hat P}{\hat \rho}\,. 
\label{eq:Helectronphonon}
\end{equation}

(v) {\it Modification of the response by $H_{\mathrm{C}}$ and $H_{\mathrm{e-p}}$.}
The function $\Xi_{jj}$ defined by Eq.~(\ref{eq:Elandph2}) 
can be written as the sum of four contributions: 
$\chi_{A}$ [the term involving ${\hat J}_{\mathrm{bl}}^{\mathrm{p}}(t)$ and 
${\hat J}_{\mathrm{bl}}^{\mathrm{p}}(0)$],  
$\chi_{B}$ [${\hat j}_{\mathrm{ph}}^{\mathrm{p}}(t)$ and 
${\hat j}_{\mathrm{ph}}^{\mathrm{p}}(0)$], 
$\chi_{C}$
[${\hat J}_{\mathrm{bl}}^{\mathrm{p}}(t)$ and 
${\hat j}_{\mathrm{ph}}^{\mathrm{p}}(0)$],  and 
$\chi_{D}$
[${\hat j}_{\mathrm{ph}}^{\mathrm{p}}(t)$ and 
${\hat J}_{\mathrm{bl}}^{\mathrm{p}}(0)$]. 
These correlators can be expressed, using perturbation theory, 
in terms of those referring to the case of $H_{\mathrm{C}}=0$ and $H_{\mathrm{e-p}}=0$: 
$\Pi_{jj}$, $\Pi_{j\rho}$, $\Pi_{\rho j}$, $\Pi_{\rho \rho}$, 
$\Lambda_{jj}$, $\Lambda_{jP}$, $\Lambda_{Pj}$ and $\Lambda_{PP}$. 
Two of the latter, $\Pi_{jj}$ and $\Lambda_{jj}$, have been defined above, 
see Eqs.~(\ref{eq:FromI3}) and (\ref{eq:Fonon2}), 
the others are defined similarly,  
for example $\Lambda_{jP}(\omega)= 
(i Na^{2}/\hbar)\int_{-\infty}^{\infty}\,{\mathrm d}t\, 
e^{i\omega t}\langle 
[{\hat j}_{\mathrm{ph}}^{\mathrm{p}}(t),{\hat P}(0)]\rangle \theta(t)$.  
Figure 1 shows the Feynman diagrams representing the correlators [in a) and b)] 
and the interaction vertices [in c)]. 
\begin{figure}[tbp]
\includegraphics{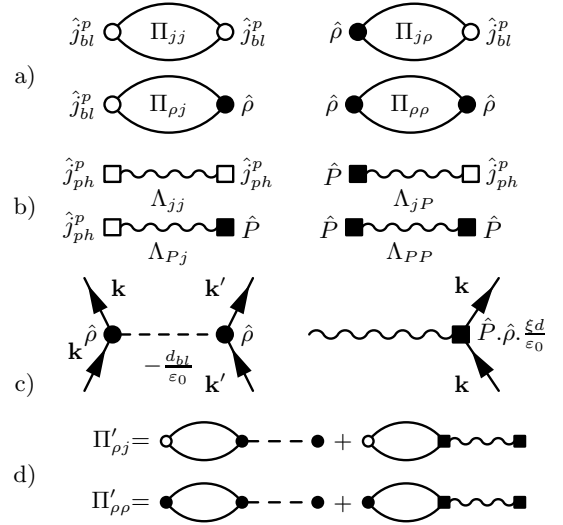}
\caption{Feynman diagrams representing the correlators defined in the text [a) and b)], 
the interaction vertices corresponding 
to $H_{C}$ of Eq.~(\ref{eq:InterlayerCoulomb}) 
and $H_{\mathrm{e-p}}$ of Eq.~(\ref{eq:Helectronphonon}) [c)], 
and two segments that appear in the series shown in Fig.~\ref{fig:diagrams} [d)]. 
}
\label{fig:segments}
\end{figure}
Also shown are the factors to be assigned to the vertices
when calculating the contributions of the diagrams. 
Figure 2 shows the series of the diagrams corresponding 
to~$\chi_{A}$, $\chi_{C}$, $\chi_{C}$ and $\chi_{D}$. 
\begin{figure*}[tbp]
\includegraphics{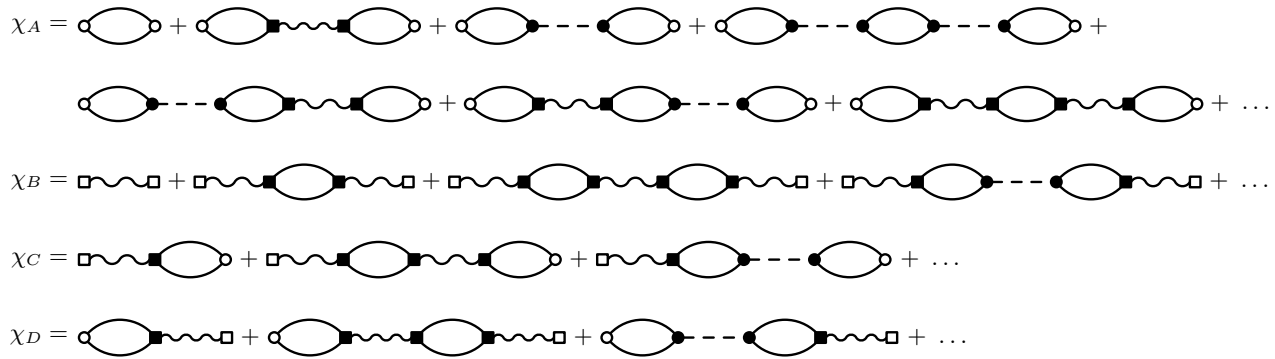}
\caption{Series of the diagrams corresponding to the quantities 
$\chi_{A}$, $\chi_{B}$, $\chi_{C}$, and $\chi_{D}$ defined in the text. 
The sum of the contributions of the diagrams is to be multiplied  
by $d^{2}_{\mathrm{bl}}/d^{2}$ in case of $\chi_{A}$ and 
by $d_{\mathrm{bl}}/d$ in case of $\chi_{C}$ and $\chi_{D}$. 
}
\label{fig:diagrams}
\end{figure*}
The series can be summed up by standard techniques\cite{Schrieffer:1988} and the results are 
$$
\left({d\over d_{\mathrm{bl}}}\right)^{2}
\chi_{A}
=\Pi_{jj}+\Pi_{j\rho}{1\over 1-\Pi'_{\rho\rho}}\Pi'_{\rho j}\,,
$$
$$
\chi_{B}=\Lambda_{jj}+
{\xi^{2}d^{2}\over \varepsilon_{0}^{2}}
\Lambda_{jP}
{\Pi_{\rho\rho}\over 1-\Pi'_{\rho\rho}}
\Lambda_{Pj}\,,
$$
$$
{d\over d_{\mathrm{bl}}}
\chi_{C}=
{\xi d\over \varepsilon_{0}}
\Pi_{j\rho}{1\over 1-\Pi'_{\rho\rho}}\Lambda_{Pj}\,,
$$
\begin{equation}
{d\over d_{\mathrm{bl}}}
\chi_{D}=
{\xi d\over \varepsilon_{0}}
\Lambda_{jP}
{1\over 1-\Pi'_{\rho\rho}}\Pi_{\rho j}\,,
\label{eq:chiABCD}
\end{equation}
where $\Pi'_{\rho\rho/\rho j}$ is the expression corresponding to the repeating segment 
of the diagrams shown in Fig.~1 (d), 
\begin{equation}
\Pi'_{\rho\rho/\rho j}={d_{\mathrm{bl}}\over\varepsilon_{0}}
\left({\xi^{2}d^{2}\over d_{\mathrm{bl}}\varepsilon_{0}}\Lambda_{PP}-1\right)\Pi_{\rho\rho/\rho j}\,. 
\label{eq:Pirhorhoorjprime}
\end{equation}

(vi) {\it Simplifications resulting from the gauge invariance.} 
The sum in the numerator on the right hand side of Eq.~(\ref{eq:Elandph1})
containing $\kappa_{\mathrm{dia-bl}}$, $\kappa_{\mathrm{dia-ph}}$ and the four correlators of Eq.~(\ref{eq:chiABCD})
can be considerably simplified by using the relations: 
$$
\Pi_{\rho j}={i\omega\varepsilon_{0}
\over d_{\mathrm{bl}}}
\chi_{\mathrm{bl}}
=-\Pi_{j\rho}\,,\,
\Pi_{\rho\rho}={\varepsilon_{0}\over d_{\mathrm{bl}}}\chi_{\mathrm{bl}}\,,
$$
\begin{equation}
\Lambda_{Pj}={i\omega\varepsilon_{0}
\over d}
\chi
=-\Lambda_{jP}\,,\,
\Lambda_{PP}={\varepsilon_{0}\over d}\chi\,,  
\end{equation}
that follow from the requirement of the charge conservation 
formulated in different gauges of the electromagnetic potentials
or equivalently from the requirement of the gauge 
invariance\cite{Chaloupka:2009:PRB,Chaloupka:2009:DP}.  
We obtain 
\begin{equation}
\sigma^{*}(\omega)=
-i\omega\varepsilon_{0}
{(d_{\mathrm{bl}}/d)\chi_{\mathrm{bl}}+\chi+(1-2\xi)\chi_{\mathrm{bl}}\chi\over 
1-\chi_{\mathrm{bl}}[(d/d_{\mathrm{bl}})\xi^{2}\chi-1]}\,.
\label{eq:sigmastar}
\end{equation}

(vii) {\it Total electric field, from $\sigma^{*}(\omega)$ to $\sigma(\omega)$}. 
The total electric field $E$ consists of the external field $E^{\mathrm{ext}}$ and 
the induced field $E^{\mathrm{ind}}$ due to the charge fluctuations described by $\rho$
and the phonon polarization described by $P$, 
$E^{\mathrm{ind}}=(d_{\mathrm{bl}}/d)(\rho/\varepsilon_{0})-(P/\varepsilon_{0})
=j/(i\omega\varepsilon_{0})$. 
We have used the continuity equation, 
$i\omega\rho=j_{\mathrm{bl}}$  
and the relation between $P$ and the total phonon current density $j_{\mathrm{p}}$, 
$-i\omega P=j_{\mathrm{p}}$, $j$ is the total current density.   
The $c$-axis conductivity $\sigma(\omega)$ defined 
by $\sigma(\omega)=j(\omega)/E(\omega)$ is equal 
to $\sigma^{*}(\omega)[E^{\mathrm{ext}}(\omega)/E(\omega)]$, i.e., 
$$
\sigma(\omega)={\sigma^{*}(\omega)\over 1-i\sigma^{*}/(\omega\varepsilon_{0})}=
$$
\begin{equation}
-i\omega\varepsilon_{0}
{(d_{\mathrm{bl}}/d)\chi_{\mathrm{bl}}+\chi+(1-2\xi)\chi_{\mathrm{bl}}\chi\over 
1+
[1-(d_{\mathrm{bl}}/d)]\chi_{\mathrm{bl}}-
\chi-
[1-2\xi+(d_{\mathrm{bl}}/d)\xi^{2}]\chi_{\mathrm{bl}}\chi}\,.
\label{eq:finalfsigmac}
\end{equation}

(viii) Considerations of {\it screening by high frequency processes} lead to a slightly modified 
formula: 
\begin{equation}
\sigma(\omega)= {\rm r.\,h.\,s.\,of\,(19)}
\left[
\varepsilon_{0}\rightarrow \varepsilon_{0}\varepsilon_{\infty}\,,\,
\chi_{\mathrm{bl}}\rightarrow {\chi_{\mathrm{bl}}\over \varepsilon_{\infty}}\,,\,
\chi\rightarrow {\chi\over \varepsilon_{\infty}}\right]\,,
\label{eq:Withepsinfty}
\end{equation}  
where $\varepsilon_{\infty}$ is the interband dielectric function.
The total dielectric function is given by
$\varepsilon(\omega)=\varepsilon_{\infty}+i\sigma(\omega)/(\omega\varepsilon_{0})$. 

It is an easy exercise to show that the final formula [(\ref{eq:finalfsigmac}) or (\ref{eq:Withepsinfty})]
is precisely the same as the one resulting from the single phonon version 
of the phenomenological multilayer model.\cite{singlephMLM}.
The important point is that is has been derived here on microscopic grounds, using perturbation theory, 
with the electron-phonon coupling constant determined by simple electrostatics. 
Within the phenomenological model, the response of a phonon to the applied field 
is governed by the eigenvector-dependent local electric field.
Here, the  eigenvector enters the equations through the coupling constant.  
Our result adds credibility to the central claim of Ref.~\onlinecite{Dubroka:2011:PRL}, 
that at $T^{\mathrm{ons}}$ 
the spectral weight of the low-energy component 
of the (local) intrabilayer conductivity $\sigma_{\mathrm{bl}}$ 
begins to increase, possibly a manifestation of precursor superconductivity.
Future studies involving the present formalism in conjunction 
with the local conductivities resulting from microscopic calculations 
such as those reported in CBM (rooted in the Fermi liquid theory) or  
those of the kind of Ref.~\onlinecite{DeNobrega:2011:PRB} 
based on a phenomenological ansatz\cite{Rice:2012:RPP} 
related to the resonating valence bond picture
are needed to achieve a further progress. 
 
This work was supported by the Ministry of Education of Czech Republic (MSM0021622410), 
by the project CEITEC - Central European Institute of Technology (CZ.1.05/1.1.00/02.0068)
from European Regional Development Fund and by the internal project MUNI/A/1047/2009. 
We gratefully acknowledge discussions with J.~Chaloupka and A.~Dubroka.


\begin{thebibliography}{99}
\bibitem{finiteq} For a finite (yet sufficiently small) ${\mathbf q}$,  
a slightly different strategy has to be adopted.  
In this case, the formula (\ref{eq:Elandph1}) 
describes the response to the total transverse field rather then the external field, 
there is no need for the correction described in the point (vii).
On the other hand, 
the coupling terms of Eqs.~(\ref{eq:InterlayerCoulomb}) and (\ref{eq:Helectronphonon}) 
are different from the text,    
since they include additional terms due to the long-range nature of the Coulomb interaction. 
For example, $H_{C}\rightarrow [1/(2\varepsilon_{0})]Na^{2}d_{\mathrm{bl}}[1-(d_{\mathrm{bl}}/d)]{\hat \rho}^{2}$.
Assuming the continuity of the current-current correlators, however, 
the same final results are obtained.
\end{thebibliography}
\end{document}